\documentclass[11pt]{eptcs}
\pdfoutput=1

\usepackage{booktabs} 
\usepackage{tikz}
\usepackage{flowchart}
\usepackage{xspace} 
\usetikzlibrary{arrows,arrows.meta,calc,decorations.pathmorphing,decorations.markings,backgrounds,decorations,decorations.pathmorphing,positioning,fit,automata,shapes,patterns,decorations.text}
\usepackage{varwidth} 
\usepackage{amsmath} 
\usepackage{amssymb} 
\usepackage{multirow} 
\usepackage{tabularx} 
\usepackage{textcomp} 

\newcommand{\Reals}{\mathbb{R}}

\newcommand{\vect}[1]{\mathbf{#1}}
\newcommand{\vx}{\vect{x}}
\newcommand{\vu}{\mathbf{u}}

\newcommand{\vy}{\vect{y}}

\newcommand{\vxdot}{\dot{\vect{x}}}

\newcommand{\domain}{\mathcal{D}}

\newcommand{\exclregionx}{1.0}
\newcommand{\exclregiony}{\pi /16}
\newcommand{\roix}{5.0}
\newcommand{\roiy}{(pi/2 - \epsilon)}

\newcommand{\simsolset}{\Phi}

\newcommand{\xpos}{x_v}
\newcommand{\ypos}{y_v}
\newcommand{\carangle}{\theta_v}
\newcommand{\xposdot}{\dot{x}_v}
\newcommand{\yposdot}{\dot{y}_v}
\newcommand{\carangledot}{\dot{\theta}_v}
\newcommand{\carspeed}{V}
\newcommand{\pathangle}{\theta_r}
\newcommand{\pathangledot}{\dot{\theta}_r}
\newcommand{\distanceerr}{d_{err}}
\newcommand{\angleerr}{\theta_{err}}
\newcommand{\distanceerrdot}{\dot{d}_{err}}
\newcommand{\angleerrdot}{\dot{\theta}_{err}}
\newcommand{\xOnPath}{x_p}
\newcommand{\yOnPath}{y_p}
\newcommand{\nncontrolfunc}{h}
\newcommand{\states}{\vx}
\newcommand{\costfnc}{J}
\newcommand{\genf}{W}

\let\OldTexttrademark\texttrademark
\renewcommand{\texttrademark}{\OldTexttrademark\xspace}%

\DeclareMathAlphabet{\mathpzc}{OT1}{pzc}{m}{it}

\newcommand{\ie}{{i.e.}\xspace}

\newcommand{\ignore}[1]{}

\newcommand{\fig}{Figure}
\newcommand{\eq}{Eq.}

\providecommand{\publicationstatus}{Invited paper in Design Automation Conference (DAC), 2018}

\title{Reasoning about Safety of Learning-Enabled Components in
Autonomous Cyber-physical Systems}

\author{Cumhur Erkan Tuncali\textsuperscript{1} \quad\ James Kapinski\textsuperscript{1} \quad Hisahiro Ito\textsuperscript{1} \quad Jyotirmoy V. Deshmukh\textsuperscript{2}
	\institute{\textsuperscript{1}Toyota Research Institute of North America, Ann Arbor, MI, USA\\
	\textsuperscript{2} University of Southern California, Los Angeles, CA, USA}
	\email{\{cumhur.tuncali, jim.kapinski, hisahiro.ito\}@toyota.com, jyotirmoy.deshmukh@usc.edu}}

\def\titlerunning{Reasoning about Safety of Learning-Enabled Components in
	Autonomous Cyber-physical Systems}
\def\authorrunning{C.E.~Tuncali, J.~Kapinski, H.~Ito \& J.V.~Deshmukh}

\begin{document}

\maketitle

\begin{abstract}

We present a simulation-based approach for generating barrier
certificate functions for safety verification of cyber-physical
systems (CPS) that contain neural network-based controllers.
A linear programming solver
is utilized to find a candidate generator function from a set of
simulation traces obtained by randomly selecting initial states for the
CPS model. 
A level set of the generator function is then selected to act as a 
barrier certificate for the system, 
meaning it demonstrates that no unsafe system states are 
reachable from a given set of initial states. 
The barrier certificate properties are verified with an SMT solver.
This approach is demonstrated
on a case study in which a Dubins car model of an autonomous vehicle
is controlled by a neural network to follow a given path.

\end{abstract}


\section{Introduction}\label{sec:intro}
Self-driving cars, unmanned aerial vehicles, and certain kinds of
robots are examples of {\em autonomous cyber-physical systems} (ACPS),
that is, physical systems controlled by software that are envisioned to
have no human operator. Remarkable success has been achieved by AI and
machine learning algorithms in solving complex tasks heretofore
thought to require human intellect. This has led to a concerted effort
to utilize AI in embedded software for ACPS applications. We observe
that the rapid advances in AI have focused on expanding the scope and
efficacy of the underlying techniques, but from a rigorous
mathematical perspective, there has been little achieved towards guaranteeing 
formal correctness of AI algorithms and their impact on overall safety
of ACPS applications in which they may be used. 

There has been a sudden upsurge in research focusing on formal
verification and testing for AI algorithms in the last two years
\cite{dutta2017output,ehlers2017formal,katz2017reluplex,huang2017safety}.
These papers focus on analyzing the AI artifacts (such as artificial
neural networks, specifically focusing on deep neural networks).  Such
analysis provides a better understanding of the robustness and safety
of the artifact itself. {\em When accompanied by environment models},
above analyses could be used to reason about the overall system safety
as well; however, such decompositional models, where the environment
assumptions are provided in a form that is easily composable with the
verification or testing algorithms for AI artifacts, are difficult to
obtain. On the other hand, approaches such as
\cite{dreossi2017compositional} take a markedly different approach;  
they perform {\em in situ} reasoning about the AI artifact in a
closed-loop model of an ACPS. In our opinion, such approaches provide
greater value by directly reasoning about the closed-loop system
safety. In this paper, we propose a method for {\em verification} of
closed-loop system models of ACPS, where the controller uses a neural
network (NN). Thus, our work goes one step further
from existing approaches as it brings mathematical rigor through
formal verification to {\em closed-loop ACPS models}.

Our key idea is to automatically learn safety invariants for the
closed-loop model. Such safety invariants can take the form of {\em
barrier certificates}. We automatically synthesize candidate barrier
certificates using simulation-guided techniques, such as those
proposed in \cite{Kapinski14,Balkan15,balkan2018underminer}. We then
verify the overall system safety by checking the validity of the
barrier certificate conditions for the candidate using the
nonlinear $\delta$-satisfiability solver dReal \cite{gao2013dreal}.

We demonstrate our technique on a simple case study of a
path-following autonomous vehicle. This vehicle uses an 
NN-based controller that was trained using reinforcement learning
(policy learning using evolutionary strategies). 
We prove that for the given kinematic models of
motion of the vehicle, it never leaves a ``safe'' region around a
given fixed path.  We caution the reader that as this is the first
attempt at verification of closed-loop ACPS models, we have not yet
applied our technique to real-world ACPS designs, and have thus not
encountered the concomitant scalability challenges. Nevertheless,
our preliminary
results are promising, as we are able to handle NN
controllers with {\em a thousand neurons} and {\em nonlinear 
activation functions}.

\section{Preliminaries}\label{sec:prelim}
 In traditional control theory, techniques such as
proportional-integral-derivative (PID) control and model predictive
control (MPC) employ {\em stateful} controllers, 
whose behaviors are defined by dynamic equations that are functions of
inputs and internal controller states.  On the other hand, control
schemes such as linear-quadratic regulator (LQR) control use {\em
stateless} controllers whose behaviors are defined by instantaneous
mappings from inputs to outputs.  The goal of this work is to analyze
systems where the stateful and stateless controllers use AI-based
algorithms.  The most popular of these are controllers based on NNs
trained using {\em reinforcement learning} approaches such as {\em
policy learning}. There are two main kinds of neural controllers, the
first of which is stateless controllers, which  employ a form of
feedforward nonlinear control. These controllers could use {\em
shallow} or {\em deep} NNs. The other kind of neural controllers are
those based on {\em recurrent neural networks} (RNNs); these employ
{\em feedback control} and are stateful.

In this work, we focus on controllers based on (stateless) feedforward
NNs, but note that the general approach outlined in this paper is
applicable to closed-loop models with RNNs as well, with the caveat
that a stateful controller will increase the query complexity of the
verification question that we frame. We will further elaborate on this
aspect later.

We consider a plant model described as follows:
\begin{eqnarray} \label{eq:plant}
\dot{\vx} & = & f_p(\vx, \vu), \\
\vy & = & g(\vx), 
\end{eqnarray}
where $\vx\in \Reals^n$ is the state, $\vu\in \Reals^m$ is the input
to the plant, $f_p$ is a locally Lipschitz-continuous vector field,
and $g:\Reals^n\rightarrow \Reals^q$ defines the plant outputs.

The NN controller is given by 
\begin{equation}
\vu = h(\vy), \label{eq:stateless_controller}
\end{equation}
where $h:\Reals^q\rightarrow \Reals^m$ is a function that maps plant
outputs to plant inputs.  We assume that $h$ performs all of the
processing required to implement the NN, including applying the
weights and activation functions that define the NN, as well as
performing any necessary preprocessing of the inputs.  We assume that
the controller is stateless and locally Lipschitz-continuous.

Composing the plant with the controller yields
\begin{eqnarray*} 
\dot{\vx} & = & f_p(\vx, h(g(\vx))),
\end{eqnarray*}
which we simplify to the following form:
\begin{eqnarray} \label{eq:composed}
\dot{\vx} & = & f(\vx).
\end{eqnarray}
Equation (\ref{eq:composed}) represents a \emph{closed-loop} model of
the system, in the sense that it is a synchronous composition of the
dynamical systems representing the plant model with the controller
model to obtain an autonomous system model (i.e., a dynamical system
with no exogenous inputs).  

\paragraph{Feedforward Neural Controller}
A feedforward neural controller is an NN that continuously maps a
multi-dimensional control input $\vy$ (of dimension $q$) to a control
output $\vu$ of dimension $m$. Following standard convention, for a
network with $M$ layers, we assume layer $1$ is the input layer, and
layer $M$ is the output layer, while layers $2,\ldots,M-1$ are the
hidden layers. The $\ell^{th}$ layer ($1<\ell\le M$) contains $d_\ell$ neurons, and the
output of the $j^{th}$ neuron in the $\ell^{th}$ layer is denoted
$v_{\ell,j}$, while the vector of outputs for the $\ell^{th}$ layer is
denoted $V_\ell$. The $\ell^{th}$ layer is parameterized by a
$d_{\ell} \times d_{\ell-1}$ weight matrix, 
where $d_1$ is the number of inputs, $W_\ell$ and a bias vector
$B_\ell$ of size $d_\ell$. $V_\ell$ is given by the expression
$\sigma_\ell(W_\ell\cdot V_{\ell -1} + B_\ell)$, where $\sigma_\ell$ is a
suitable activation function applied component-wise for the $\ell^{th}$ layer.

We note that previous work from the formal methods community focused
on NNs with ReLU (rectified linear unit) activation functions. A ReLU
with input $v$ essentially computes $\max(v,0)$; that is, it is
piecewise linear in the input, which makes it amenable to analysis by
SMT solvers equipped with linear theories. We do not impose any such
restriction on activation functions, as we reduce the verification
questions to nonlinear queries over real numbers, which can be
analyzed by dReal -- a nonlinear SMT solver based on interval
constraint propagation. dReal is capable of handling Type 2
computable functions, which are essentially real functions that can be
numerically approximated. These include polynomials, trigonometric
functions, and exponentials \cite{Gao2012}.  Thus, we allow activation
functions such as the sigmoid function $\sigma (v) = 1/(1+e^{-v})$ and
the hyperbolic tangent function $\sigma(v) = \tanh(v)$ (implemented in
MATLAB\textsuperscript{\textregistered} as the {\em tansig} function, which has a faster
implementation than the $\tanh$ function).

\subsection{Safety Verification with Strict Barrier Certificates}

A barrier certificate is an inductive invariant function for
contin-uous-time dynamical systems \cite{Prajna04,Prajna2005}. We
assume that we are given an autonomous dynamical system described by
(\ref{eq:composed}), a set of possible initial states $X_0$, and a set
of unsafe states $U$. Then, we define the barrier certificate as
follows.

\paragraph{Strict Barrier Certificate}
\label{def:sbc}
A barrier certificate is a differentiable function $B$ from the set of
states of the dynamical system to the set of reals. Let $\nabla B$
denote the gradient of $B(\vx)$, \ie, $\nabla B = [\frac{\partial
B}{\partial x_1}\ \frac{\partial B}{\partial x_2} \ldots\
\frac{\partial B}{\partial x_n}]$.  Then, $B(\vx)$ is called a strict
barrier certificate for a dynamical system of the form specified in
Eq.~\eqref{eq:composed}, if it satisfies the following conditions:
\begin{eqnarray*} 
	(1)& \forall \vx \in X_0:& B(\vx) \le 0, \\
	(2)& \forall \vx \in U:& B(\vx) > 0,\\
	(3)& \forall \vx :& (B(\vx) = 0) \implies (\nabla B)^T\cdot f(\vx) < 0 .
\end{eqnarray*}

We observe that the existence of a suitable barrier certificate
demonstrates that along any system trajectory with the initial state
in $X_0$, a state in $U$ is not reachable (in finite or infinite
time).  Thus, a barrier certificate provides a powerful unbounded-time
safety certificate of the system.

\newpage
\section{Solution Overview}\label{sec:overview}
We present a method to perform verification of safety properties for
CPSs that contain NN components.  Our approach closely follows the
simulation-based barrier certificate strategy described in
\cite{Kapinski14}. The key idea in this approach is to define a
barrier certificate as a level set of a {\em generator function}
$\genf(\vx)$, i.e. the barrier certificate $B(\vx)$ is the function
$\genf(\vx) - \ell$ for some $\ell \in \Reals^{> 0}$. The generator
function $\genf(\vx)$ is assumed to be a {\em positive} function that
decreases along the system trajectories. We assume that $\genf(\vx)$ is
specified using suitable templates, such as Sum-of-Squares
polynomials, where the coefficients of the monomial terms are to be
determined. 

The method starts by performing a collection of simulations to
generate a set of linear constraints that specify the positivity of
the candidate generator function, and that it decreases along system
trajectories. We then check condition (3) from
Definition~\ref{def:sbc}; note that we can do this as $\nabla B$ =
$\nabla \genf$, since $\ell$ is a constant. We check this condition using an
SMT solver. The SMT solver either produces a counterexample (CEX) 
that results in an updated candidate generator function, or it returns
UNSAT, which certifies that the candidate is sound.  Finally, we use
the generator function to find the appropriate value of $\ell$ that
separates the initial condition set from the unsafe set, and thus acts
as a barrier certificate for the system.  There are certain nuances in
each of these steps that we now describe below.

The flowchart in Figure \ref{fig:NNVerificationFlowhart} illustrates
the process.  We first create a collection of linear constraints, 
as described above, using
results from simulations $\simsolset_s$. A linear program (LP) is
solved to obtain a solution that satisfies the constraints.  The LP
solution corresponds to a candidate generator function $\genf(\vx)$.
\begin{figure}[htbp]
	\centering
	\begin{tikzpicture}[>=stealth',scale=1]
\tikzstyle{smalltext}=[font=\fontsize{7}{7}\selectfont]
\tikzstyle{bigtext}=[font=\fontsize{9}{9}\selectfont]
\tikzstyle{block}=[draw,fill=white,process,minimum height=2em,text width=3em,smalltext,align=center]
\tikzstyle{data}=[draw,fill=white,trapezium,trapezium left angle=70,trapezium right angle=-70,minimum height=2em,text width=5.7em,inner sep=1pt,smalltext,align=center]


\node[data,text width=1.5cm] (init) 
     {Seed Traces $\simsolset_s$};
\node[draw,circle,inner sep=2pt,minimum height=1em,text width=1em,bigtext,below of=init,node distance=10mm,align=center] (sum) 
     {$\displaystyle\uplus$};
\node[data,left of=sum,node distance=25mm,text width=0.8cm] (pairs) 
     {Traces $\simsolset_f$};
\node[draw,block,below of=sum,node distance=10mm] (solvelp) 
     {Solve LP$^{(1)}$};
\node[data,below of=solvelp,node distance=12mm,text width=1.85cm] (lyap)
     {Candidate Generator Function $\genf$};
\node[draw,block,below of=lyap,node distance=12mm,text width=1.5cm] (SMT1) 
     {SMT Solver: Check (\ref{eq:barrierquery1})};
\node[draw,smalltext,diamond,aspect=2,text width=4.5em,below of=SMT1,node distance=15mm,inner sep=0pt,text height=1em,align=center] (decision1) 
     {UNSAT?};
     
\node[data,left of=SMT1,node distance=25mm,text width=0.5cm] (cex) 
{CEX};
\node[draw,block,below of=pairs,node distance=15mm,text width=1.1cm] (simulate) 
{Simulate};

\node[draw,block,right of=init,node distance=40mm] (complevelset) 
     {Compute Level set};
\node[data,below of=complevelset,node distance=12mm,text width=.6cm] (levelset)
     {Level set};
\node[draw,block,below of=levelset,node distance=12mm,text width=1.5cm] (checkL) 
     {SMT Solver: Check (\ref{eq:barrierquery2}) \& (\ref{eq:barrierquery3})};
\node[draw,smalltext,diamond,aspect=2,text width=4.5em,below of=checkL,node distance=15mm,inner sep=0pt,text height=1em,align=center] (decision2) 
     {UNSAT?};
\node[draw,rounded corners, below of= decision2, node distance=20mm,text width=2.5cm] (stop) 
     {Halt:\\System is Safe};  

\draw[->] (init)     to (sum);
\draw[->] (sum)      to (solvelp);
\draw[->] (pairs)    to (sum);
\draw[->] (solvelp)  to (lyap);
\draw[->] (lyap)     to (SMT1);
\draw[->] (SMT1)     to (decision1);
\draw[->] (decision1.west) -| node[above,very near start] {\textsc{No$^{(2)}$}} ($(cex.south)$) ;
\draw[->] (cex) to (simulate);
\draw[->] (simulate) to (pairs);
\draw[->] (decision1.east) -| node[above,very near start] {\textsc{Yes}} ($(complevelset.west)+(-1.0,0.0)$) -- ($(complevelset.west)$); 
\draw[->] (complevelset)  to (levelset);
\draw[->] (levelset)  to (checkL);
\draw[->] (checkL)  to (decision2);
\draw[->] (decision2) to node[left] {\textsc{Yes}} (stop);
\draw[->] (decision2.east) -| node[above,very near start] {\textsc{No$^{(3)}$}} ($(complevelset.east)+(1.5,0.0)$) -- ($(complevelset.east)$);

\node[text width=10cm, below of=cex] at (1.55, -7) (footnote)
{ $^{(1,2,3)}$ If the LP is infeasible or if the maximum number of iterations to find a candidate generator function or a levelset is reached, the algorithm terminates with no conclusions.};


\end{tikzpicture}
	\caption{Procedure to verify safety property for NN-based system.}\label{fig:NNVerificationFlowhart}
\end{figure}

Next, an SMT solver is used to check the following property over the
domain of interest: 
\begin{eqnarray} \label{eq:barrierquery1} 
\exists \vx \in \domain:\ (\vx \not\in X_0) \wedge ((\nabla \genf)^T\cdot f(\vx))~\geq -\gamma.  
\end{eqnarray} 
The above query is UNSAT when for all $\vx \in \domain\setminus X_0$,
the condition $((\nabla \genf)^T\cdot f(\vx)) < 0$ holds, which means the
condition $((\nabla B)^T\cdot f(\vx)) < 0$ holds (because $B(\vx) =
\genf(\vx) - \ell$). Let $L = \{ \vx \mid \genf(\vx) - \ell \le 0 \}$.  Note
that the boundary of $L$ (denoted $\partial L$) is the set where
$B(\vx) = 0$.  We later explain how we choose $\ell$ such that $X_0
\subset L$ , and $L \subset \domain$. In other words, we check the
condition $((\nabla B)^T \cdot f(\vx)) < 0$ over a set that is a
superset of $\partial L$, or the set $B(\vx) = 0$. Thus, the above
condition being UNSAT, guarantees that condition (3) in
Def.~\ref{def:sbc} holds.  For our experiments, we use
$\gamma=10^{-6}$.

If the SMT solver returns SAT, then a corresponding CEX is returned in
the form of an $\vx^*$ such that $\nabla B(\vx^*)^T\cdot f(\vx^*) \geq
-\gamma$.  This CEX is then used to generate new linear constraints,
based on a simulation, $\simsolset_f$, obtained by using the CEX as an
initial condition, and then another LP is solved to produce an updated
candidate generator function. This iterative process continues until
the SMT solver returns UNSAT.

Next, we try to compute the level set size $\ell$ of $\genf(\vx)$ such that the set
$L$ (i.e. $\{\vx \mid B(\vx) < 0\}$) contains the initial condition
set $X_0$ and does not intersect with the unsafe set $U$.  The methods
available to select an appropriate $\ell$ value will depend on the
class of the chosen generator function $\genf$ and the geometry of sets
$X_0$ and $U$.  In the examples provided in the subsequent section,
$\genf$ is a quadratic function, $X_0$ is a rectangle, and $U$ is a
disjunction of halfspaces.  For this case, the set $L$ is an
ellipsoid, and $\ell$ can be selected to be any value that satisfies
the following:
\begin{itemize}
\item Each vertex of $X_0$ lies within $L$;
\item Each halfspace defining $U$ is disjoint from the ellipsoid $L$.
\end{itemize}

Once the level set size $\ell$ is selected, a pair of additional SMT
queries is performed to check whether $X_0 \subset L$ and $L \cap U =
\emptyset$.  As with (\ref{eq:barrierquery1}), we check the
satisfiability of the negation of these conditions with the SMT
solver:
\begin{eqnarray} \label{eq:barrierquery2}
\exists \vx \in X_0, \vx \notin L, \\ \label{eq:barrierquery3}
\exists \vx \in L, \vx \in U, 
\end{eqnarray}
which will return UNSAT if the desired property holds. If either of
these queries returns SAT, then the level set $L$ does not satisfy the
desired properties, and a new $\ell$ value should be selected. We can
do this efficiently by performing a binary search on a feasible range
of $\ell$ values until the SMT solver returns UNSAT for the queries 
\ref{eq:barrierquery2} and \ref{eq:barrierquery3}.

If the final pair of queries returns UNSAT, then the procedure halts,
and the function $B(\vx) = \genf(\vx) - \ell$ is proven to be a barrier
certificate for the system, meaning that the system is proven to be
safe.  In the next section, we present an example that demonstrates
the above method to prove safety for an ACPS with an NN controller.

We note that in formal proofs of unsatisfiability, it is important to
pay attention to the interpretation of mathematical functions and constants. 
For example, in the context of the verification approach shown in
Fig. \ref{fig:NNVerificationFlowhart}, 
the mechanisms used to generate traces $\simsolset_s$ and $\simsolset_f$ 
and the SMT solver should ideally have the same interpretation of the system
dynamics.
For our implementation, we use MATLAB\textsuperscript{\textregistered} to produce $\simsolset_s$ and $\simsolset_f$ 
and dReal to address the SMT queries, but
MATLAB\textsuperscript{\textregistered} and dReal may have slightly different interpretations of, for example,
the exponential functions found in the activation functions and the constants
that define the NN weights.
We sidestep this issue by assuming the following: a.) 
the MATLAB\textsuperscript{\textregistered} interpretation of the system dynamics
is only an approximation used to generate candidate 
generator functions, and b.) the system dynamics in the 
``deployed'' implementation, including 
the weights and functions used
to define the NN controller, are the same used for the dReal queries.

\section{Case Study}\label{sec:casestudies}
\begin{figure}[htbp]
	\centering
	\begin{tikzpicture}[auto, node distance=5cm,>=latex',scale=0.9, every node/.style={scale=0.9}]

\tikzset{%
  every neuron/.style={
    circle,
    draw,
    minimum size=0.5cm
  },
  neuron missing/.style={
    draw=none, 
    scale=1,
    text height=0.15cm,
    execute at begin node=\color{black}$\vdots$
  },
}

\newcommand\XS{2.4}
\newcommand\XO{3.4}

\tikzstyle{block} = [draw, fill=green!20, rectangle, 
    minimum height=2em, minimum width=6em,align=center,execute at begin node=\setlength{\baselineskip}{1.5em}]
     \tikzstyle{texts} = [minimum height=4em, minimum width=6em,align=center,execute at begin node=\setlength{\baselineskip}{1.5em}]
\tikzstyle{input} = [coordinate]
\tikzstyle{output} = [coordinate]


    
    \node [block,anchor=center, fill=blue!20,minimum width = 4.4cm, minimum height = 3.3cm] at (0,0) (NN) {};
    
\node [coordinate,text width=5em,label={[align=center]Neural Network (Controller)}] at ($(NN.north)$) (NNlabel) {};

   \node [block, fill=white,minimum width = 1.4cm, minimum height = 1.4cm] at ($(NN.170)+(-2.5,0)$) (reference){} ;

	\node [coordinate,text width=5em,label={[align=center]Ref. Trajectory}] at ($(reference.north)$) (Reflabel) {};

 \draw [->] ($(reference.west)+(0.15,-0.55)$) -- ($(reference.west)+(0.15,0.55)$) ;
 \draw [->] ($(reference.west)+(0.1,-0.45)$) -- ($(reference.west)+(1.3,-0.45)$) ;
 \draw [-,blue] ($(reference.west)+(0.15,-0.45)$) -- ($(reference.west)+(0.5,-0.1)$) ;
 \draw [-,blue] ($(reference.west)+(0.5,-0.1)$) -- ($(reference.west)+(0.8,0.0)$) ;
 \draw [-,blue] ($(reference.west)+(0.8,0.0)$) -- ($(reference.west)+(1,0.2)$) ;
 \draw [-,blue] ($(reference.west)+(1,0.2)$) -- ($(reference.west)+(1.0,0.4)$) ;
 \draw [-,blue] ($(reference.west)+(1.0,0.4)$) -- ($(reference.west)+(1.1,0.5)$) ;

\node [every neuron] (Hidden1) at ([xshift=\XS cm, yshift=1.2cm]NN.west) {};
\node [every neuron] (Hidden2) at ([xshift=\XS cm, yshift=0.6cm]NN.west) {};
\node [every neuron] (Hidden3) at ([xshift=\XS cm, yshift=0.0cm]NN.west) {};
\node [neuron missing] (Hidden4) at ([xshift=\XS cm, yshift=-0.7cm]NN.west) {};
\node [every neuron] (Hidden5) at ([xshift=\XS cm, yshift=-1.2cm]NN.west) {};

\node [every neuron] (Output1) at ([xshift=\XO cm, yshift=0.6cm]NN.west) {};

    \node [block, rotate=90, text height = .5em](preprocessing) at ([xshift=1.5em]NN.west) {\textsc{Preprocessing}} ;
    
\node [block,below of=NN, fill=red!20,text width=8em,yshift=2.2cm] (dynamics) {$\vxdot=f_p(\vx,\vu)$};
\node [coordinate,text width=5em,label={[align=center]Car Model}] at ($(dynamics.north)$) (Dynamicslabel) {};

\node [] at ($(preprocessing.south)+(0,0.5)$) (distnode) {};
\node [] at ($(preprocessing.south)+(0,-0.1)$) (thetanode) {};

 \draw [->] (distnode.center) -- node[above,pos=0.5] {$\distanceerr$} ($(distnode)+(0.75,0)$) -- ($(Hidden1.west)$) ;
 \draw [->] ($(distnode)+(0.75,0)$) -- ($(Hidden2.west)$) ;
 \draw [->] ($(distnode)+(0.75,0)$) -- ($(Hidden3.west)$) ;
 \draw [->] ($(distnode)+(0.75,0)$) -- ($(Hidden5.west)$) ;
 
 \draw [->] (thetanode.center) -- node[above,pos=0.5] {$\angleerr$}  ($(thetanode)+(0.75,0)$) -- ($(Hidden1.west)$) ;
 \draw [->] ($(thetanode)+(0.75,0)$) -- ($(Hidden2.west)$) ;
 \draw [->] ($(thetanode)+(0.75,0)$) -- ($(Hidden3.west)$) ;
 \draw [->] ($(thetanode)+(0.75,0)$) -- ($(Hidden5.west)$) ;

 \draw [->] ($(Hidden1.east)$) -- ($(Output1.west)$) ;
 \draw [->] ($(Hidden2.east)$) -- ($(Output1.west)$) ;
 \draw [->] ($(Hidden3.east)$) -- ($(Output1.west)$) ;
 \draw [->] ($(Hidden5.east)$) -- ($(Output1.west)$) ;
           
 \draw [->] ($(Output1.east)$) --  node[above,pos=0.5] {$u$} ($(Output1.east)+(0.5,0)$) ;           
       
 \draw [very thick,->] ($(NN.east)$) -- ($(NN.east)+(1.0,0)$) |- node[above,pos=0.9] {$u$} ($(dynamics.east)$) ;   
 \draw [very thick,->] ($(dynamics.west)$) -- node[above,pos=0.1]  {$\vx$} ($(dynamics.west)+(-2.0,0)$) |- ($(NN.190)$) ; 
 \draw [very thick,->] ($(reference.east)$) -- ($(NN.170)$) ;     
       
\end{tikzpicture}
	\caption{Closed-loop simulation setup.}\label{fig:nn_in_loop}
\end{figure}
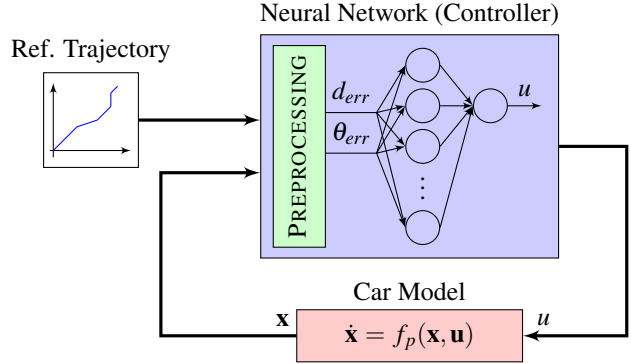

In this section, we present a case study that we use to evaluate our
verification approach.  We consider a Dubins car, where an NN
controller is used to track a given path.  An overview of the
closed-loop system is provided in \fig\ \ref{fig:nn_in_loop}.  We
first describe the system dynamics.  Then, we describe the technique
we used to develop an NN controller.  Finally, we demonstrate the
proposed verification approach on the case study.

\subsection{System Dynamics}
\subsubsection{Vehicle Model}

We first describe the kinematic model of the Dubins car.  \fig\
\ref{fig:path_following} (a) illustrates the notation used for the
position ($\xpos, \ypos$) and the orientation ($\carangle$) of the
vehicle on the 2-D ($x,y$) plane.  The orientation ($\carangle$) is
defined as the clockwise angle with respect to the positive $y$-axis.
The magnitude of the longitudinal velocity is denoted by $\carspeed$.

The Dubins car model uses the following differential equations to
represent the car dynamics: 
\begin{align}
	\xposdot =& V\sin{\carangle}\label{eq:xpos_dot}\\
	\yposdot =& V\cos{\carangle}\label{eq:ypos_dot}\\
	\carangledot =& u\label{eq:input_u}
\end{align}
In the above kinematic model, $u$ is the turn rate control, which we
will refer to as \textit{steering} control. For our experiments, we
assume that car velocity, $V$, is constant.

\subsubsection{Path Following}
\begin{figure}[t]
	\begin{center}
		\begin{tabular}{cc}
			\begin{tikzpicture}[scale=1.0]
    \draw[->,semithick] (0,-.75) -- (0,3.5);
    \draw (-0.2,3.15) node {$y$};
    \draw[->,semithick] (-.25,-0.5) -- (4,-0.5);
    \draw (3.7,-0.7) node {$x$};
    
    
     \draw [draw=blue,rotate=-40,semithick] (-0.3,2) rectangle ++(0.6,1.5);
     \draw[dashed,rotate=-40,draw=blue] (0,0) -- (0,2.75);
     \draw[-{>[scale=1.5]},rotate=-40,draw=blue,thick] (0,2.75) -- (0,4.5);
     \draw[dashed,draw=gray] (1.7677, 2.1066) -- (1.7677, -0.55);
     \draw[dashed,draw=gray] (1.7677, 2.1066) -- (-0.05, 2.1066);
     \node[black] at (-0.3, 2.1)  {$\ypos$};
     \node[black] at (1.85, -0.7)  {$\xpos$};
     \node[blue] at (2.5, 3.5)  {$V$};
     
     \draw[->,black] (0,1) arc (90:50:1);
     \node[black] at (0.4,1.15)  {$\carangle$};
\end{tikzpicture}\hspace{0in}
			&\begin{tikzpicture}[scale=1.0]

    \draw[->,semithick] (0,-0.75) -- (0,3.5);
    \draw (-0.2,3.15) node {$y$};
    \draw[->,semithick] (-.25,-0.5) -- (4,-0.5);
    \draw (3.7,-0.7) node {$x$};
    
    \draw[<-,red,thick] (1,3.5) arc (0:-60:3.5);
    \draw[->,red] (0,1.8) arc (90:62:1.0);
    \node[red] at (0.3,2)  {$\pathangle$};
    
    \coordinate[](A) at (0,0.8);
    \draw[dashed,draw=red,rotate around={-30:(A)}] (0,-0.75) -- (0,3.8);
     \draw [draw=blue,rotate=-70,semithick] (-0.3,1.5) rectangle ++(0.6,1.5);
     \draw[dashed,rotate=-70,draw=blue] (0,-1) -- (0,2.25);
     \draw[->,rotate=-70,draw=blue,thick] (0,2.25) -- (0,4);
     
     \draw[->,blue] (0,0.4) arc (90:20:0.4);
     \node[blue] at (0.4,0.5)  {$\carangle$};
     
     \draw[<->,semithick] (2.05,0.77) -- (0.5,1.66);
     \node[black] at (1.5,1.5)  {$\distanceerr$};
     \draw[<-,black] (0.15,1.0) arc (70:33:1.6);
     \node[black] at (0.85,1)  {$\angleerr$};
     
     \draw[dashed, gray] (0.5,1.66) -- (1.5,2.5);
     \node[gray] at (2.25, 2.6) {$(x_p, y_p)$};
\end{tikzpicture}\hspace{0in}\\
			(a) Dubins car model.&(b) Path following errors.
		\end{tabular}
		\caption{Notation used in system dynamics.}
		\label{fig:path_following}
	\end{center}
\end{figure}
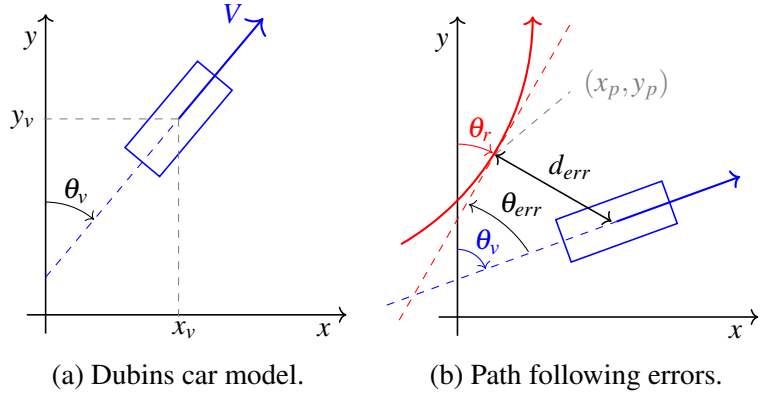

For any given vehicle state and target path, we compute the distance
error and angle error of the vehicle with respect to the target path.
\fig\ \ref{fig:path_following} (b) illustrates the computation of
distance and angle errors.  The solid red curve represents a section
of the target path.  The distance error, which is denoted as
$\distanceerr$, is defined as the shortest distance from the vehicle
coordinates to the target path.  On the target path, the closest point
to the vehicle is denoted as $(\xOnPath, \yOnPath)$.  The angle error,
which is denoted as $\angleerr$, is the angle between the vehicle
orientation ($\carangle$) and the orientation of the tangent line to
the target path at the point  $(\xOnPath, \yOnPath)$.  Hence, if the
angle of the tangent line is defined as $\pathangle$, the angle error
is defined as follows:
\begin{equation}
	\angleerr = \pathangle - \carangle. \label{eq:angleerr}
\end{equation}
The distance error, $\distanceerr$, is taken as negative when the angle error is negative and its absolute value is smaller than $\pi$, which is when the vehicle is on the right of the target path, as shown in \fig\ \ref{fig:path_following} (b) and positive when the vehicle is on the left of the target path.

\subsubsection{Error Dynamics}
Based on the vehicle dynamics and the equations defining the path
following error, we use the error dynamics to define a system model as
follows.  For simplicity, we consider the target path as a straight
line with a constant orientation $\pathangle$.  Assume that the target
path starts at the coordinate $(0, 0)$, which is a reasonable
assumption given that the origin of the coordinate system may be
shifted so that the target path starts at the origin.  Then, we rotate
the coordinate system by $\frac{\pi}{2}-\pathangle$ radians in a
clockwise direction.  In the rotated coordinate system, the target
trajectory becomes the $x$-axis, and the rotated $y$ coordinate of the
vehicle is taken as the distance error.  This means that, when the
vehicle is on the right side of the trajectory, the distance error
will be negative and vice versa when it is on the left.  The following
is the distance error:
\begin{align}
	\distanceerr = -\xpos\sin{\big(\frac{\pi}{2}-\pathangle\big)} + \ypos\cos{\big(\frac{\pi}{2}-\pathangle\big)} .\label{eq:distance_err}
\end{align}
Hence, following from \eq\ \ref{eq:xpos_dot}, \ref{eq:ypos_dot},
\ref{eq:angleerr} and \ref{eq:distance_err}, the time derivative of
$\distanceerr$ can be computed as follows:
\begin{align*}
	\distanceerrdot &= -\xposdot\cos{(\pathangle)} + \yposdot\sin{(\pathangle)}\\
	 &= -V\sin{(\carangle)}\cos{(\pathangle)} + V\cos{(\carangle)}\sin{(\pathangle)} \\
	 &= -V\sin{(\pathangle - \angleerr)}\cos{(\pathangle)} + V\cos{(\pathangle - \angleerr)}\sin{(\pathangle)}.
\end{align*}
Furthermore, following from \eq\ \ref{eq:input_u} and
\ref{eq:angleerr} and the fact that the path angle is constant, the
time derivative of $\angleerr$ is given by:
\begin{equation}
	\angleerrdot\ = \pathangledot - \carangledot\ = -u .
\end{equation}

\subsubsection{Closed Loop System Dynamics}
Considering the NN controller as a function, $\nncontrolfunc$, mapping
its inputs $\distanceerr$ and $\angleerr$ to its output $u$, where $u$
is the input to the plant, the closed loop system dynamics can be
defined as follows, where $\mathbf{\states}$ denotes the system state
vector:
\begin{align*}
	\mathbf{\states} &= [\distanceerr\ \angleerr]^T \\
	\dot{\mathbf{\states}} &= 
	f_p(\vx, \vu)\\
	&=
	\begin{bmatrix}
		-V\sin{(\pathangle - \angleerr)}\cos{(\pathangle)} + V\cos{(\pathangle - \angleerr)}\sin{(\pathangle)}\\
		-u
	\end{bmatrix}\\
	\vy &= g(\vx) = [\distanceerr\ \angleerr]^T \\
	u &= h (\vy).
\end{align*}

\subsection{Learning a Controller}

To learn an NN controller, we first select the structure of the
network.  We elect to use a feedforward NN with one hidden layer and
$N_{h}$ neurons in the hidden layer.  The NN takes distance and angle
errors ($\distanceerr, \angleerr$) as inputs, and it outputs steering
control $u$.  Hence, the input layer accepts two inputs, and the
output layer contains one neuron.  An NN with $N_{h}$ neurons in the
hidden layer with the structure we have selected has $(1\times N_{h})
+ (N_{h}\times 2)$ weight parameters and $N_{h} + 1$ bias parameters.
Hence the total number of parameters (including weights and bias
values) is $4 N_{h} + 1$.  We used \textit{tansig} for all activation
functions.  The implementation of the NN is done in
MATLAB\textsuperscript{\textregistered} . 

\begin{figure}[htbp]
	\begin{center}
		\setlength\tabcolsep{1.4pt}
		\begin{tabular}{cc}
			\includegraphics[width=.30\linewidth]{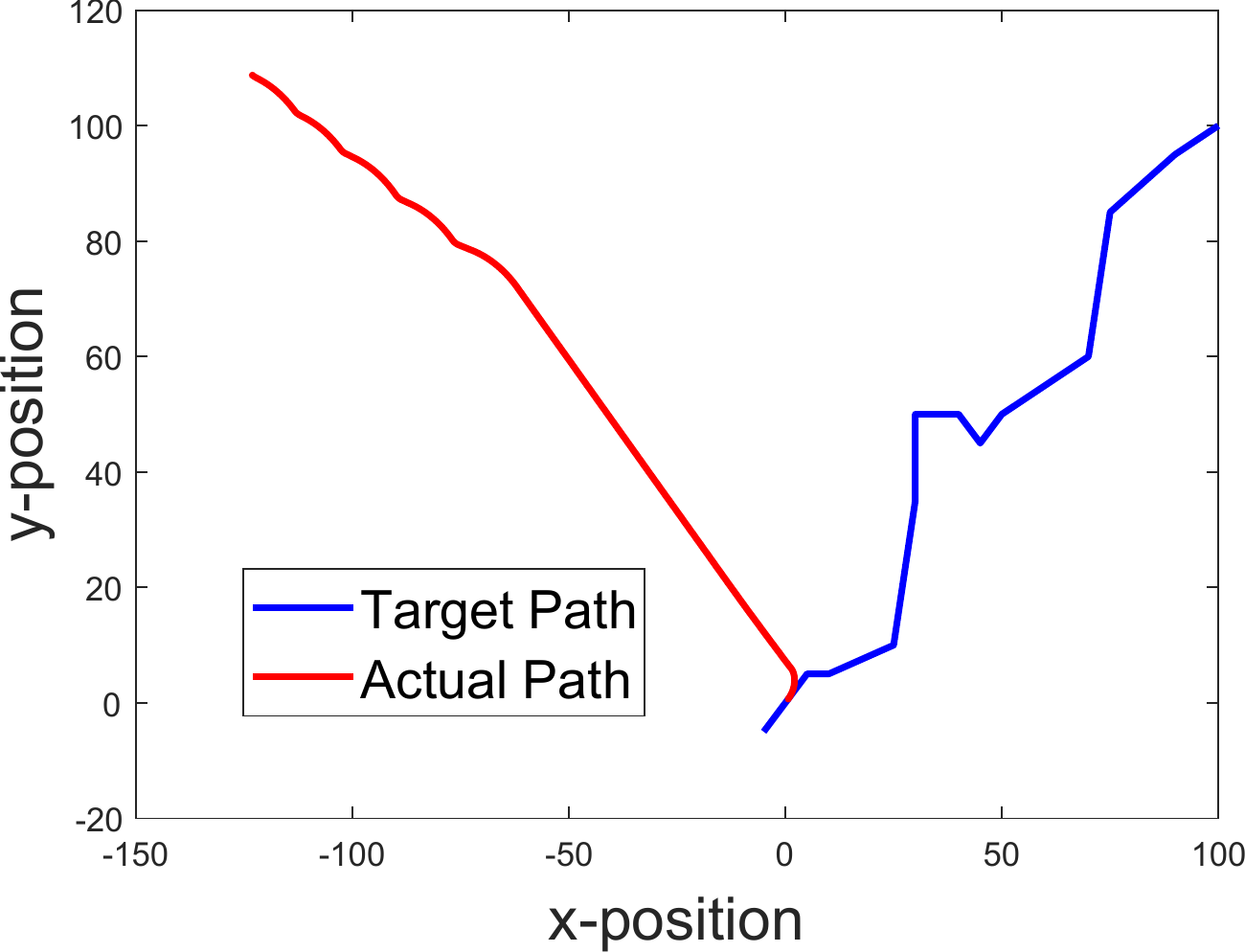}\hspace{0in}
			&\includegraphics[width=.30\linewidth]{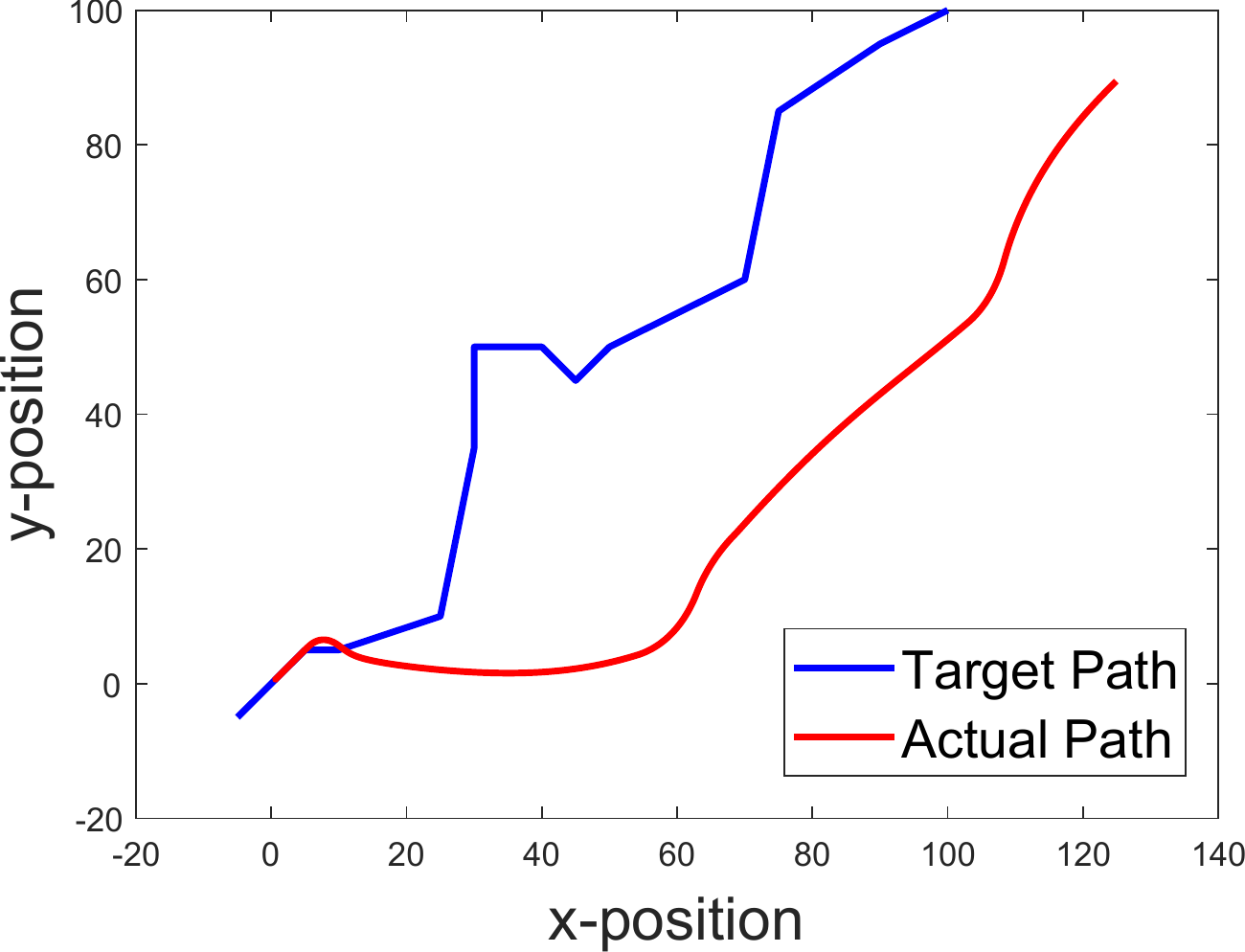}\hspace{0in}\\
			(a) With random intial weights&(b) At iteration 5\vspace{0.12in}\\
			\includegraphics[width=.30\linewidth]{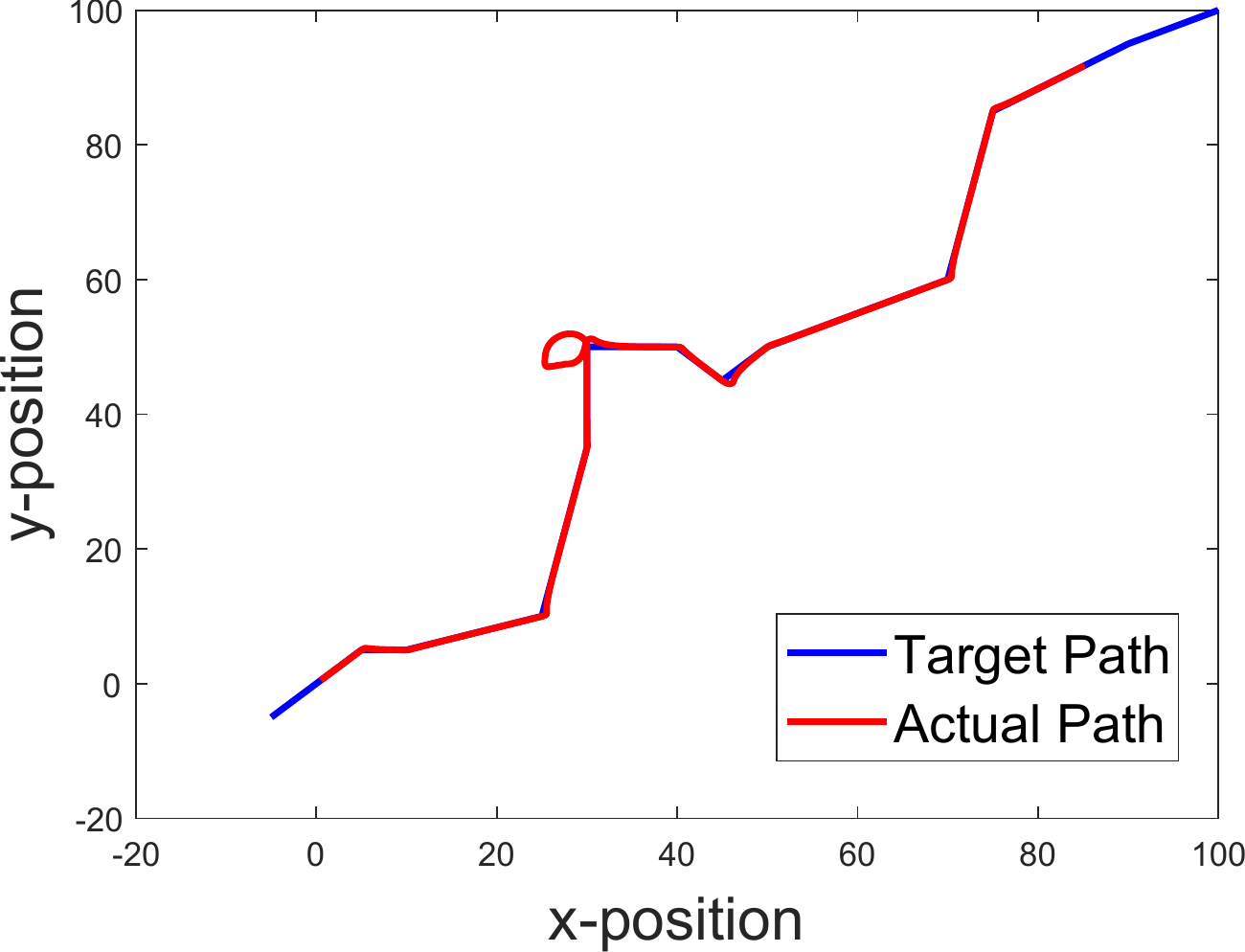}\hspace{0in}
			&\includegraphics[width=.30\linewidth]{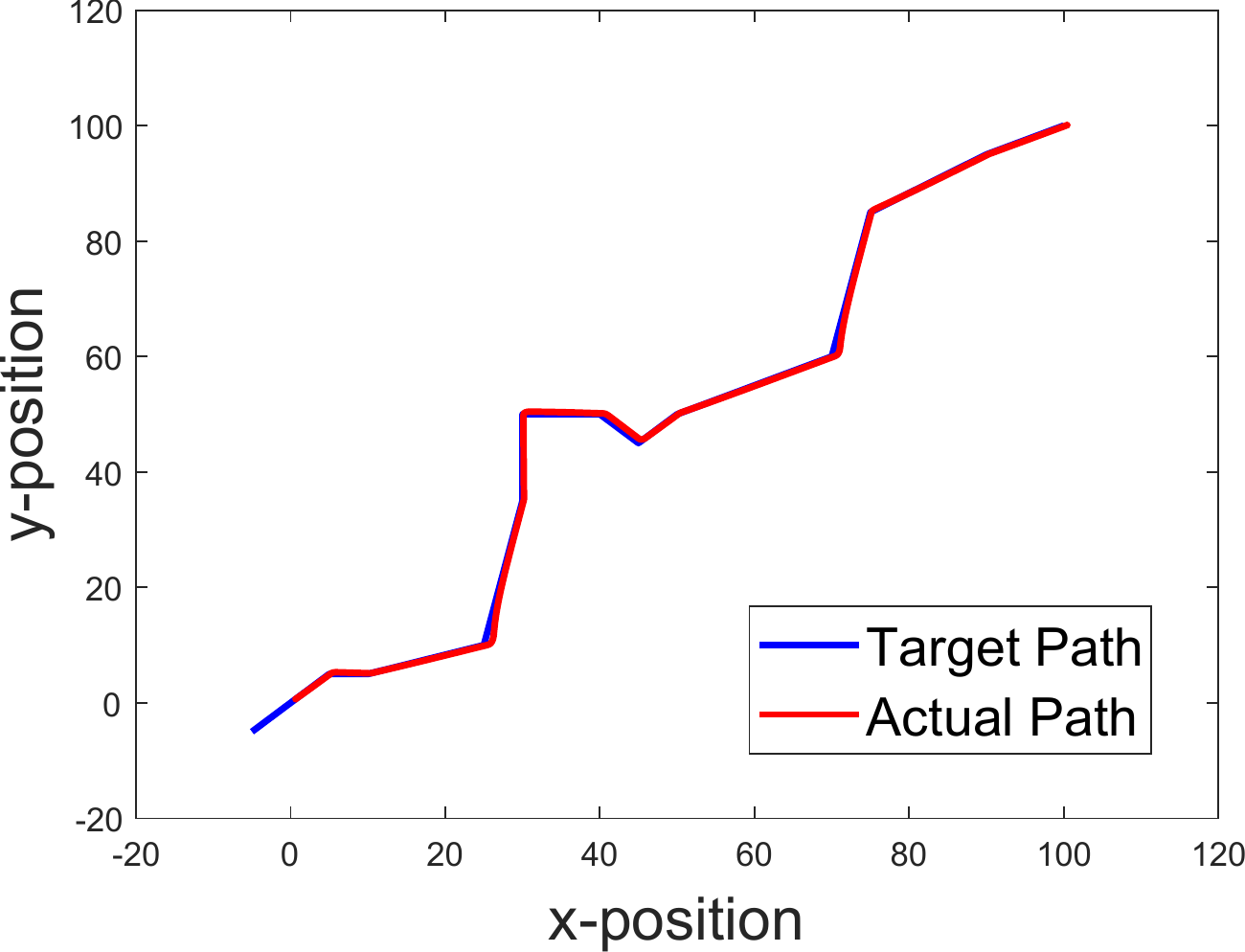}\\
			(c) At iteration 25&(d) At the end of the training\\
		\end{tabular}
		\caption{Evolution of the NN controller during the policy search.}
		\label{fig:cmaes_evolution}
	\end{center}
\end{figure}

By starting with a random set of NN parameters, we performed direct
policy search variant of reinforcement learning using a CMA-ES
algorithm \cite{hansen2001completely, igel2003neuroevolution} to find
an optimal set of parameters (weights and biases) for the NN
controller.  For the direct policy search, we used the blue
(piecewise-linear) path shown in \fig\ \ref{fig:cmaes_evolution} as
the target path on the x-y plane.  The CMA-ES algorithm is used to
optimize the NN parameters with the goal of minimizing the path
following error.  From a discrete-time simulation of the system with a
controller that is using the parameters that come from CMA-ES, we
compute a corresponding cost using the following cost function:
\begin{align*}
\costfnc =& \sum_{k\in\{0, ..., N\}}\Big(100{\distanceerr}_k^2 + 10^5{\angleerr}_k^2 + 100u_k^2\Big) \\
&+ 10^3|(x_{end}, y_{end})-({\xpos}_{N}, {\ypos}_{N})|^2 .
\end{align*}
Note that $N$ represents the number of discrete time steps in the
simulation.  The subscripted terms ${\distanceerr}_k, {\angleerr}_k$
and $u_k$ represent the corresponding values of ${\distanceerr},
{\angleerr}$ and $u$, respectively, at the time step $k$ of the
simulation.  The last term in the cost function computes the error
related to the Euclidean distance between the end point of the the
target path, $(x_{end}, y_{end})$, and the final position of the
vehicle, $({\xpos}_{N}, {\ypos}_{N})$, in the simulation.

\fig\ \ref{fig:cmaes_evolution} illustrates some sample simulation
traces from the evolution of an NN controller with $10$ neurons in the
hidden layer, using the policy search based on CMA-ES optimization
with a maximum number of $50$ iterations and a population size of
$152$.

The final parameter values arrived at by the CMA-ES algorithm are used
as fixed weights and biases for the NN controller.  Note that, after
the training phase was completed,  we validated the performance of the
controller informally by observing behaviors for a set of random
reference trajectories, and we observed reasonable performance from
the system.

\subsection{Verification Results}

We applied the approach described in Section \ref{sec:overview} to our
case study, for a number of different versions of the NN controller
described above.  Each version of the system we consider contains a
different number of neurons in the hidden layer.  By evaluating our
technique on this suite of systems, we demonstrate how well the method
scales with the size of the NN.

For our evaluations, we assume that the target path for the controller
is a straight line. For each verification, $X_0$ is given by the
rectangular area defined by the diagonal corners $(-\exclregionx,
-\exclregiony)$ and $(\exclregionx, \exclregiony)$, and $U$ is the
complement (outside) of the rectangle described by the diagonal
corners $(-\roix, -\roiy)$ and $(\roix, \roiy)$. The domain of
interest for the barrier search is defined as $\domain=(X_0 \cup U)'$,
where $S'$ is the complement of set $S$.

\begin{figure*}[htbp]
	\centering
	\includegraphics[width=\textwidth]{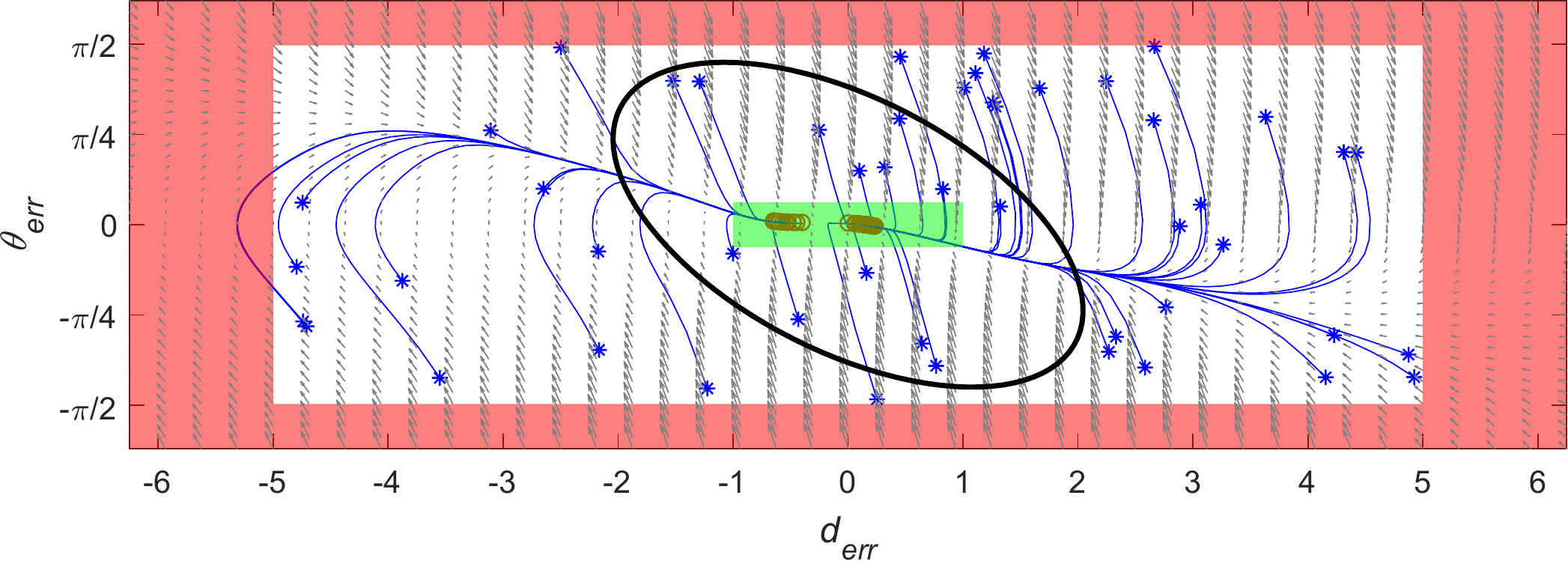}
	\caption{Phase portrait with sample system trajectories, $X_0$, $U$ and a barrier certificate level set.}
	\label{fig:barrier_certificate}
\end{figure*}

Table \ref{table:results} presents the experimental results.  Each row
of the table reports the time taken and number of iterations for each
step of the procedure described in
Fig.~\ref{fig:NNVerificationFlowhart} applied to different versions of
the system shown in Fig.~\ref{fig:nn_in_loop} (where the versions
differ in the complexity of the neural network-based controller used).
The numbers shown correspond to average values over $30$ experiments;
each experiment uses a unique seed to generate the initial simulations
used to produce $\simsolset_s$ in Fig.
\ref{fig:NNVerificationFlowhart}.  The first column of the table
indicates how many neurons are present in the hidden layer of the NN.
The second column indicates the time spent to find a generator
function (i.e., the time taken to complete the iterations of the first
loop in Fig.  \ref{fig:NNVerificationFlowhart}).  The third column
indicates the average number of iterations needed to find a generator
function.  Each iteration consists of Solve LP and SMT Solver Check
(\ref{eq:barrierquery1}) operations as shown in
Fig.~\ref{fig:NNVerificationFlowhart}.  The third and fourth columns
indicate the average time spent in each execution of Solve LP and SMT
Solver Check (\ref{eq:barrierquery1}) operations, respectively.  The
fifth column indicates the total amount of time spent in the
operations given in Fig. \ref{fig:NNVerificationFlowhart} and not
captured in the previous columns.  The last column indicates the total
time.  Experimental results show that our approach scales well with
the increasing number of neurons in the controller.  We note that
dReal uses heuristics to perform branch and prune operations, and
although our experimental results show that it is generally able to
solve long queries quickly, in the worst-case, SMT solutions can be
costly.  This occasional poor performance is exemplified in some of
our experiments (e.g., the 300 and 500 neuron cases).  We refer the
reader to \cite{Gao2012,gao2013dreal} for a more detailed
computational complexity analysis for dReal.

\begin{table}[htb] 
\centering  

{\small
\begin{tabular}{@{\extracolsep{\fill}}r crrr l r r}
\toprule
\multicolumn{1}{r}{Number}    & \multicolumn{4}{c}{Computing Generator} & & Time Spent& Total \\
\cline{2-5} 
\multicolumn{1}{r}{of}   & Avg. Num. & \multicolumn{3}{c}{Time Spent (s.)} & &  in Other& Time \\
\cline{3-5}
\multicolumn{1}{r}{Neurons}     & Iterations  & LP & Query    & Total
& &  Steps (s.) & (s.) \\
\midrule
10   & 3.0 & 1.1 & 4.25 & 43  & & 12 & 55 \\
20   & 1.8 & 1.1 & 2.6 & 21  &  & 11 & 32 \\
40   & 1.7 & 1.2 & 5.3 & 26  & & 14 & 40 \\
50   & 1.5 & 1.6 & 4.8 & 35  & & 14 & 49 \\
70   & 2.8 & 1.8 & 15.6 & 106 & & 16 & 122 \\
80   & 1.2 & 1.7 & 4.3 & 28  & & 15 & 43 \\
90   & 1.0 & 2.0 & 4.7 & 27  & & 16 & 43 \\
100  & 1.7 & 1.1 & 4.1 & 21  & & 14 & 35 \\
300  & 1.7 & 1.8 & 379.8 & 698 & & 48 & 746 \\
500  & 1.3 & 1.9 & 379.4 & 536 & & 107 & 643 \\
700  & 1.0 & 2.0 & 19.1 & 41  & & 35 & 76 \\
1000 & 1.0 & 2.0 & 50.4 & 74  & & 79 & 153 \\
\bottomrule
\end{tabular} 
}


\caption{Timing analysis on safety verification for various versions of the system illustrated in Fig. \ref{fig:nn_in_loop}. 
\label{table:results}}
\end{table}

\fig\ \ref{fig:barrier_certificate} illustrates the results of 
verification for one of the cases captured in Table
\ref{table:results}.  The lateral axis of the figure represents the
position error (i.e., the $\distanceerr$ state), and the vertical axis
represents the angle error ($\angleerr$).  The initial condition set
$X_0$ is shown in green, and the unsafe set $U$ is shown in red.  The
simulation trajectories  $\simsolset_s$ are shown in blue; initial
conditions for each trajectory are marked with an ($\ast$) and end
points are marked with a ($\circ$).  The sample space for the initial
states is the $\domain$ region.  The ellipsoid between the $X_0$ and
$U$ sets in \fig\ \ref{fig:barrier_certificate} is a level set of a
generator function found using our approach; the barrier
properties (\ref{eq:barrierquery1}), (\ref{eq:barrierquery2}), and
(\ref{eq:barrierquery3}) are all determined to be UNSAT by the dReal
SMT solver \cite{gao2013dreal}.  Hence, the ellipsoid is a barrier
certificate for the system, which means that the system is safe.

\section{Conclusion}\label{sec:conclusion}
In this paper, we present a technique to reason about safety of a
closed-loop control system using a learning-enabled controller. 
In particular, we focus on feedforward artificial neural network-based controllers. 
The key idea of our approach is to reduce the
safety verification problem to the identification of a barrier
certificate candidate, using simulations of the closed-loop system,
and then perform {\em a posteriori} verification of the synthesized barrier
certificate. The final verification step is performed using a
nonlinear SMT solver, which permits our approach to handle neural
networks with arbitrary nonlinear activation functions. We demonstrate
the feasibility of our technique on a simple closed-loop model of a
path-following ground vehicle. Future work will focus on improving the
scalability of our technique and investigating stateful controllers
based on recurrent neural networks. We will also investigate
algorithms to simultaneously train the neural network while satisfying
safety guarantees.

\bibliographystyle{eptcs}
\bibliography{nnsafety-bibliography}

\end{document}